\documentclass[12pt]{article}
\usepackage{amsfonts,amssymb,amsmath}
\usepackage[dvips]{epsfig}
\begin{document}
\title{\bf Perfect routing of quantum information in regular cavity QED networks}
\author{ N. Behzadi $^{a}$
\thanks{E-mail:n.behzadi@tabrizu.ac.ir} ,
S. Kazemi Rudsary $^{b}$  ,
B. Ahansaz Salmasi$^{b}$
\\ $^a${\small Research Institute for Fundamental Sciences,}
\\ $^b$ {\small Department of Theoretical Physics
and Astrophysics, Faculty of Physics,}
\\ {\small University of
Tabriz, Tabriz 51666-16471, Iran.}} \maketitle
\begin{abstract}
We introduce a scheme for perfect routing of quantum states and entanglement in regular cavity QED networks. The couplings between the cavities are quasi-uniform and each cavity is doped with a two-level atom. Quasi-uniform couplings leads the system to evolve in invariant subspaces. Combination the evolutions of the system in its invariant subspaces with quite simple local operations on atoms in the networks, gives the perfect routing of quantum states and entanglement through the network. To provide the protocol be robust due to decoherence arisen from photon loss, the field mode of the cavities are only virtually excited.
\end{abstract}

\section{Introduction}
Realizing any effective protocol in the filed of quantum communication and distributed quantum computing depends on reliable transferring of a typical quantum state form one point to another throughout an efficient quantum communication channel. Implementing all of these protocols in the realm of quantum mechanics, needs to many body interacting quantum systems like spin chains \cite{bose} or cavity quantum electrodynamics (QED) systems which each of them contains an emitter like atom \cite{cirac}.

In many protocols of quantum state transfer based on linear spin chains, it is clear that the efficiency of the protocols depends seriously on the engineering of coupling strengths between the spins such that the perfect state transfer on spin chains with uniform coupling is possible only for chains with two and three spins. For this reason, several attempts have been done to engineer the couplings in such a way that the natural dynamics achieves state transfer perfectly [3-9] or with a reliable fidelity [10, 11]. Recently, for achieving state transmission with minimal engineering of couplings, application of external local magnetic field on the system has been effective in this way [12, 13]. The necessity of engineering of couplings, which becomes more complicated in perfect routing of a quantum state on an arbitrary path in a regular network with arbitrary spatial dimension, can be removed by taking quasi-uniform couplings $(\pm)$ which leads to obtain perfect state transfer in small regions of a spin network and tailing these regions to obtain perfect state transfer in the whole system \cite{pet, kar}.

On the other hand, communication channels with cavity QED structure have several practical superiorities in comparison to those one constructed by spins. In spin chains, single spin addressing is difficult because the spatial separation between neighboring spins is very small \cite{li}. Thus the control over the couplings between the spins or over individual spins is very hard to achieve. While the coupled cavities, which each of them contains an atom,
have the advantage of easily addressing individual cavities with optical lasers. Furthermore,
the interaction of a cavity and an atom, can be engineered in such way that
the atom trapped in the cavity can have relatively long-lived energy levels which is suitable
for encoding of quantum information \cite{uti}.

In this work, we introduce a protocol for perfect quantum routing on a regular network of cavities with arbitrary dimension, which each of cavities contains a two-level atom. A quantum states can be stored on an atom and transfer from that atom to the other one throughout photon hopping between nearest neighbor cavities. Quasi-uniform engineering of the interactions between cavities leads to have small invariant atom-cavity subspaces with perfect state transfer property as quantum routing unit. Combination the natural dynamics in invariant subspaces along with the tailing these dynamics with simple local operations leads to transfer quantum information from any sender to any receiver and vice versa in the network. This protocol can also work for entanglement transfer throughout the network.

From the practical points of view, we investigate that the process of perfect routing of quantum information, in this way, can be established in two distinct regimes. In one of them, the field mode
of each cavity exhibits resonance interaction with the atom. In this case,
the field mode can be extremely populated which in turn, leads to decay of photons to the environment quickly and
therefore lose of coherency of the system \cite{dav, fon}. But, when the field mode is highly detuned
with the atomic transition frequency, the establishment of perfect state transfer can be satisfied without
populating the field mode in each cavities. Consequently, the perfect quantum routing protocol
is susceptible to decoherence from photon loss and thus, the efficient
decoherence rate of the cavities is left to be an unimportant problem.

\section{One-dimensional prototype}
We consider, at first, routing of quantum states in one-dimension by using system of 3N+1 two-level atom trapped in single mode 3N+1 cavities arranged in such a way that depicted in Fig. 1(a). The Hamiltonian for this arrangement is as
\begin{eqnarray}
\hat{H}=\hbar\Omega_{c} \sum_{i=1}^{3N+1}\hat{a}_{i}^{\dagger}\hat{a}_{i}+\hbar\Omega_{e} \sum_{i=1}^{3N+1}\hat{\sigma}_{i+}\hat{\sigma}_{i-}+\hbar G\sum_{i=1}^{3N+1}(\hat{a}_{i}\hat{\sigma}_{i+}+\hat{a}_{i}^{\dagger}\hat{\sigma}_{i-})\nonumber
\\&&\hspace{-100mm}+\hbar\sum_{\{k,l\}\in E} J_{k,l}(\hat{a}_{k}^{\dagger}\hat{a}_{l}+\hat{a}_{k}\hat{a}_{l}^{\dagger}),
\end{eqnarray}
where $\hat{a}_{i}^{\dagger}$ and $\hat{a}_{i}$ are the creation and annihilation operators for the field mode of $i$th
cavity with frequency $\Omega_{c}$ and $\hat{\sigma}_{i+}$ and $\hat{\sigma}_{i-}$ denote the Pauli rising and lowering operators for the atom with transition frequency $\Omega_{a}$ trapped in the $i$th cavity. $G$ is the strength of atom-cavity coupling due to Jaynes-Cummings interaction \cite{jay} and $J_{kl}$ is the strength of coupling
between $k$th and $l$th cavities arisen from photon hopping between them \cite{har} and $E$ is the set of edges of the graph depicted in Fig. 1(a), corresponding to the coupling between the cavities. This Hamiltonian is excitation preserving, i.e.
\begin{eqnarray}
[\hat{H}, \hat{N}]=0,
\end{eqnarray}
where $\hat{N}$ is the number operator defined as
\begin{eqnarray}
\hat{N}=\sum_{i=1}^{3N+1}(\hat{a}_{i}^{\dagger}\hat{a}_{i}+\hat{\sigma}_{i+}\hat{\sigma}_{i-}).
\end{eqnarray}
Therefore, the Hamiltonian in Eq. 1  does not evolve the state without excitation. We consider Hamiltonian in the (6N+2)-dimensional single excitation invariant subspace whose standard basis are $\{|j\rangle\}$ with $j=1, 2, ..., 6N+2$ for which, odd $j$s corresponding to the excitation of field modes of cavities and even $j$s corresponding to the excitation of two-level atoms. It is assumed that $\hbar=1$, $\Delta \equiv\Omega_{c} - \Omega_{a}$ as detuning parameter and $J_{3n,3n+1}=-J$ for all $n=1, 2,...,N$ otherwise $J_{k,l}=J$. We introduce the other set of basis in the single excitation subspace, in terms of standard basis as
\begin{eqnarray}
\begin{array}{c}
|c_{n}\rangle:=|6n-5\rangle ,\ |a_{n}\rangle:=|6n-4\rangle ,\\\\
|c_{n}^{\pm}\rangle:=\frac{1}{\sqrt{2}} (|6n-3\rangle \pm |6n-1\rangle ) ,\ |a_{n}^{\pm}\rangle:=\frac{1}{\sqrt{2}} (|6n-2\rangle \pm |6n\rangle ) ,
\end{array}
\end{eqnarray}
where denoting that the inequality $6n-s\leq6N+2$ should be hold for all $n=1,2,...,N+1$ and $s=0,1,2,...,5$.
Hence, by choosing the set of basis in (4), the Hamiltonian is left with a direct sum structure as
\begin{equation}
\ \hat{H}=\bigoplus_{n=1}^{N+1}\hat{H}_{n},
\end{equation}
where their respective invariant subspaces can be regarded as
\begin{equation}
\begin{array}{c}
  \mathcal{H}_{1}= span\{|c_{1}\rangle,|a_{1}\rangle,|c_{1}^{+}\rangle,|a_{1}^{+}\rangle\}, \\\\
  \mathcal{H}_{n+1}= span\{|c_{n}^{-}\rangle,|a_{n}^{-}\rangle,|c_{n+1}\rangle,|a_{n+1}\rangle,|c_{n+1}^{+}\rangle,|a_{n+1}^{+}\rangle\}, n=1,2,...,N-1, \\\\
  \mathcal{H}_{N+1}= span\{|c_{N}^{-}\rangle,|a_{N}^{-}\rangle,|c_{N+1}\rangle,|a_{N+1}\rangle\}.
\end{array}
\end{equation}
In this basis, the subsystem $\hat{H}_{1}$ has the same structure as the subsystem $\hat{H}_{N+1}$ and whose matrix representation is as
\begin{equation}
\ \hat{H}_{1} = \left(
          \begin{array}{cccc}
          \Omega_{c} & G & J\sqrt{2}  & 0 \\
          G & \Omega_{c}-\Delta & 0 & 0 \\
          J\sqrt{2}  & 0 & \Omega_{c} & G \\
          0 & 0 & G & \Omega_{c}-\Delta \\
          \end{array}
          \right),
\end{equation}
which is similar to a system composed of two single mode cavities which each of them is doped with a two-level atom, and coupled to each other by a coupling strength $J\sqrt{2}$. Also, each of subsystem $\hat{H}_{2}$, $\hat{H}_{3}$, ...,$\hat{H}_{N}$ has the same structure as the other and the matrix form of one of them such as $H_{2}$, is as follows
\begin{equation}
\ \hat{H}_{2} = \left(
          \begin{array}{cccccc}
          \Omega_{c} & G & J\sqrt{2} & 0 & 0 & 0\\
          G & \Omega_{c}-\Delta & 0 & 0 & 0 & 0\\
          J\sqrt{2} & 0 & \Omega_{c} & G & J\sqrt{2} & 0\\
          0 & 0 & G & \Omega_{c}-\Delta & 0 & 0\\
          0 & 0 & J\sqrt{2} & 0 & \Omega_{c} & G \\
          0 & 0 & 0 & 0 & G & \Omega_{c}-\Delta\\
          \end{array}
          \right).
\end{equation}
It is evident that $\hat{H}_{2}$ is similar to the Hamiltonian of a system composed of three single mode cavities, arranged on a line, which each of them contains a two level atom and coupled to each other by a coupling strength $J\sqrt{2}$. The achievement of perfect transfer of quantum state $|\psi\rangle=\alpha|0\rangle+\beta|1\rangle$ with $|\alpha|^2+|\beta|^2=1$, from the first atom to the $(3N+1)$th atom, depends on the achievement of perfect state transfer in each subsystems $\hat{H}_{1}$, $\hat{H}_{2}$, ..., $\hat{H}_{N+1}$ as units of perfect quantum routing. Therefore, we should only analyze the perfect state transfer process in subsystems $\hat{H}_{1}$ and $\hat{H}_{2}$ because the remainders are similar to these ones. If we assume that the atom in the first cavity is only excited, i.e. at $t=0$ the system is prepared to be in the state $|a_{1}\rangle$, time evolution process leads to the following expression, i.e.
\begin{equation}
\hat{U}(t)|a_{1}\rangle=u_{\hat{H}_{1},1}(t) |c_{1}\rangle+u_{\hat{H}_{1},2}(t) |a_{1}\rangle +u_{\hat{H}_{1},3}(t) |c_{1}^{+}\rangle +u_{\hat{H}_{1},4}(t) |a_{1}^{+}\rangle,
\end{equation}
where $\hat{U}(t)=e^{-i\hat{H}t}$ is the time evolution unitary operator and $u_{\hat{H}_{1},i}(t)$s with $i=1,2,3,4$, as they have been brought in the appendix, satisfy the following relation:
\begin{equation}
|u_{\hat{H}_{1},1}(t)|^2+|u_{\hat{H}_{1},2}(t)|^2 +|u_{\hat{H}_{1},3}(t)|^2+|u_{\hat{H}_{1},4}(t)|^2=1,
\end{equation}
which is consequence of the Eq. 2 and indicates that dynamical time evolution of the system is restricted to the related subspace $\mathcal{H}_{1}$ with average number of photons $F_{\hat{H}_{1}}=|u_{\hat{H}_{1},1}(t)|^2 +|u_{\hat{H}_{1},3}(t)|^2$. Perfect state transfer process in the subspace $\mathcal{H}_{1}$ ensures that the state $|a_{1}\rangle$ should be transferred to the entangled state $|a_{1}^{+}\rangle$, i.e.
\begin{equation}
u_{\hat{H}_{1},4}(t_{\hat{H}_{1}})= u_{\hat{H}_{1},2}(0)=1,
\end{equation}
where $t_{\hat{H}_{1}}$ is the relative transfer time. This situation
is satisfied under some circumstances which depend on the amounts of the parameters $G$, $J$ and $\Delta$ as shown in Fig. 2. Now, after transfer time $t_{\hat{H}_{1}}$, if we apply local operation $\hat{V}=\prod_{n=1}^{N}\hat{\sigma}_{Z3n}$ which is effective only on first control part in this time, the state $|a_{1}^{+}\rangle$ transforms to the state $|a_{1}^{-}\rangle$ which belongs to the subspace $\mathcal{H}_{2}$. By the dynamics, the state $|a_{1}^{-}\rangle$ evolves as
\begin{equation}
\hat{U}(t)|a_{1}^{-}\rangle=u_{\hat{H}_{2},1}(t) |c_{1}^{-}\rangle+u_{\hat{H}_{2},2}(t) |a_{1}^{-}\rangle +u_{\hat{H}_{2},3}(t) |c_{2}\rangle +u_{\hat{H}_{2},4}(t) |a_{2}\rangle+u_{\hat{H}_{2},5}(t) |c_{2}^{+}\rangle+u_{\hat{H}_{2},6}(t) |a_{2}^{+}\rangle,
\end{equation}
where, as the previous case, the following equality holds for $u_{\hat{H}_{2},i}(t)$s as they have been given in the appendix.
\begin{equation}
|u_{\hat{H}_{2},1}(t)|^{2}+|u_{\hat{H}_{2},2}(t)|^{2}+|u_{\hat{H}_{2},3}(t)|^{2}+|u_{\hat{H}_{2},4}(t)|^{2}+|u_{\hat{H}_{2},5}(t)|^{2}+|u_{\hat{H}_{2},6}(t)|^{2}=1,
\end{equation}
where $F_{\hat{H}_{2}}:=|u_{\hat{H}_{2},1}(t)|^2 +|u_{\hat{H}_{2},3}(t)|^2+|u_{\hat{H}_{2},5}(t)|^2$, is the average number of photons for the field mode of cavities in the subspace $\mathcal{H}_{2}$. But, transmission the state $|a_{1}^{-}\rangle$ to the state $|a_{2}^{+}\rangle$ perfectly, depends on satisfying the following equation:
\begin{equation}
u_{\hat{H}_{2},6}(t_{\hat{H}_{2}})= u_{\hat{H}_{2},2}(0)=1,
\end{equation}
which holds as shown in Fig. 3. After transfer time $t_{\hat{H}_{2}}$, performing local operation $\hat{V}=\prod_{n=1}^{N}\hat{\sigma}_{Z3n}$ gives out the state $|a_{2}^{-}\rangle$ which lies in the subspace $\mathcal{H}_{3}$. The natural time evolution process gives perfectly the state $|a_{3}^{+}\rangle$ after transfer time $t_{\hat{H}_{3}}=t_{\hat{H}_{2}}$ and applying the local operation gives the state $|a_{3}^{-}\rangle$. This process can continue until to receive the state $|a_{N+1}\rangle$ which means that the quantum state $|\psi\rangle=\alpha|0\rangle+\beta|1\rangle$ transferred from the first atom to the $(3N+1)th$ atom perfectly with transfer time $T=2t_{\hat{H}_{1}}+(N-1)t_{\hat{H}_{2}}$ as depicted symbolically in Fig. 1(b). It should be noted that this method also works for entanglement transfer along the network.
\section{Extension to three-dimensional structure}
We develop the scheme for perfect quantum routing discussed in the previous section to three-dimensional structure. We start to describe the method by introducing following Hamiltonian as
\begin{eqnarray}
  &&\hspace{-10mm}\hat{H}_{\mu}=\hbar\Omega_{c} \sum_{i=0}^{3}(\hat{a}_{\mu-\nu_{i}}^{\dagger}\hat{a}_{\mu-\nu_{i}}+\hat{a}_{\mu_{i}}^{\dagger}\hat{a}_{\mu_{i}})+\hbar\Omega_{a} \sum_{i=0}^{3}(\hat{\sigma}_{+(\mu-\nu_{i})}\hat{\sigma}_{-(\mu-\nu_{i})}+\hat{\sigma}_{+\mu_{i}}\hat{\sigma}_{-\mu_{i}})\nonumber\\ &&\hspace{-5mm}+\hbar G\sum_{i=0}^{3}(\hat{a}_{\mu-\nu_{i}}\hat{\sigma}_{+(\mu-\nu_{i})}+\hat{a}_{\mu-\nu_{i}}^{\dagger}\hat{\sigma}_{-(\mu-\nu_{i})}) +\hbar G\sum_{i=0}^{3}(\hat{a}_{\mu_{i}}\hat{\sigma}_{+\mu_{i}}+\hat{a}_{\mu_{i}}^{\dagger}\hat{\sigma}_{-\mu_{i}})\nonumber\\ &&\hspace{25mm}+\hbar\sum_{i,j=0}^{3} M_{ij}(\hat{a}_{\mu_{i}}^{\dagger}\hat{a}_{\mu-\nu_{j}}+\hat{a}_{\mu_{i}}\hat{a}_{\mu-\nu_{j}}^{\dagger}),
\end{eqnarray}
where $M_{ij}$s are elements of the following matrix:
\begin{eqnarray}
M=J\left(
    \begin{array}{cccc}
      1 & 1 & 1 & 1 \\
      1 & 1 & -1 & -1 \\
      1 & -1 & 1 & -1 \\
      1 & -1 & -1 & 1 \\
    \end{array}
  \right).
\end{eqnarray}
This Hamiltonian corresponds to the set of cavities which each of them contains a two-level atom with atom-cavity and cavity-cavity couplings denoted by $G$ and $J$ respectively, as shown in Fig. 4. As the previous section, we consider the single excitation subspace with standard basis denoted by: $\{|\mu-\nu_{i}^{b}\rangle,|\mu_{i}^{b}\rangle\}$ with $b\in\{c,a\}$ and $i=0,1,2,3$, in which $c$ and $a$ stand for denoting cavities and atoms. The basis $|\mu_{i}^{b}\rangle$s are devoted to the atoms and cavities inside the frame and $|\mu-\nu_{i}^{b}\rangle$s represent the related basis for those ones outside the frame as denoted in Fig. 4. Also in this way, the other set of basis defined in terms of the standard basis, can be regarded as $\{|\mu-\nu_{i}^{b}\rangle,|\xi_{\mu i}^{b}\rangle\}$ where
\begin{eqnarray}
\begin{array}{c}
|\xi_{\mu0}^{c}\rangle := \frac{1}{2} (|\mu_{0}^{c}\rangle + |\mu_{1}^{c}\rangle +|\mu_{2}^{c}\rangle + |\mu_{3}^{c}\rangle ) ,\quad |\xi_{\mu0}^{a}\rangle := \frac{1}{2} (|\mu_{0}^{a}\rangle + |\mu_{1}^{a}\rangle +|\mu_{2}^{a}\rangle + |\mu_{3}^{a}\rangle ), \\\\
|\xi_{\mu1}^{c}\rangle := \frac{1}{2} (|\mu_{0}^{c}\rangle + |\mu_{1}^{c}\rangle -|\mu_{2}^{c}\rangle - |\mu_{3}^{c}\rangle ) ,\quad |\xi_{\mu1}^{a}\rangle := \frac{1}{2} (|\mu_{0}^{a}\rangle + |\mu_{1}^{a}\rangle -|\mu_{2}^{a}\rangle - |\mu_{3}^{a}\rangle ), \\\\
|\xi_{\mu2}^{c}\rangle := \frac{1}{2} (|\mu_{0}^{c}\rangle - |\mu_{1}^{c}\rangle +|\mu_{2}^{c}\rangle - |\mu_{3}^{c}\rangle ) ,\quad |\xi_{\mu2}^{a}\rangle := \frac{1}{2} (|\mu_{0}^{a}\rangle - |\mu_{1}^{a}\rangle +|\mu_{2}^{a}\rangle - |\mu_{3}^{a}\rangle ), \\\\
|\xi_{\mu3}^{c}\rangle := \frac{1}{2} (|\mu_{0}^{c}\rangle - |\mu_{1}^{c}\rangle -|\mu_{2}^{c}\rangle + |\mu_{3}^{c}\rangle ) ,\quad |\xi_{\mu3}^{a}\rangle := \frac{1}{2} (|\mu_{0}^{a}\rangle - |\mu_{1}^{a}\rangle -|\mu_{2}^{a}\rangle + |\mu_{3}^{a}\rangle ). \\\\
\end{array}
\end{eqnarray}
Therefore in these basis, the Hamiltonian in Eq. 15, is left with a direct sum structure as
\begin{equation}
\hat{H}_{\mu} = \bigoplus_{i=0}^{3}\hat{H}_{\mu i},
\end{equation}
where the relative subspace of each $\hat{H}_{\mu i}$ is given by
\begin{equation}
\mathcal{H}_{\mu i}= span\{|\mu-\nu_{i}^{c}\rangle,|\mu-\nu_{i}^{a}\rangle,|\xi_{\mu i}^{c}\rangle,|\xi_{\mu i}^{a}\rangle\},
\end{equation}
where each of $\hat{H}_{\mu i}$ with $i=0,1,2,3$, has the same structure as the others. For example:
\begin{equation}
\hat{H}_{\mu0} = \left(
          \begin{array}{cccc}
          \Omega_{c} & G & 2 J & 0 \\
          G & (\Omega_{c}-\Delta) & 0 & 0 \\
          2 J & 0 & \Omega_{c} & G \\
          0 & 0 & G & (\Omega_{c}-\Delta) \\
          \end{array}
          \right),
\end{equation}
which is similar to the Hamiltonian (7) in the previous section except that $J$ has been replaced by $J\sqrt{2}$. Let's consider that at $t=0$, the atom in the $\mu-\nu_{0}$ cavity be excited, i.e. the system is prepared to be in the state $|\mu-\nu_{0}^{a}\rangle$. As the previous cases, the dynamics of the system evolves this state in the subspace $\mathcal{\hat{H}}_{\mu0}$ as follows
\begin{equation}
\hat{U}(t)|\mu-\nu_{0}^{a}\rangle=u_{\hat{H}_{\mu_{0}},1}(t) |\mu-\nu_{0}^{c}\rangle+u_{\hat{H}_{\mu_{0}},2}(t) |\mu-\nu_{0}^{a}\rangle +u_{\hat{H}_{\mu_{0}},3}(t) |\xi_{\mu0}^{c}\rangle +u_{\hat{H}_{\mu_{0}},4}(t) |\xi_{\mu0}^{a}\rangle,
\end{equation}
where $\hat{U}(t)=e^{-i\hat{H}_{\mu}t}$ and $u_{\hat{H}_{\mu_{0}},i}(t)$s with $i=1,2,3,4$, are the same as $u_{\hat{H}_{1},i}(t)$s in Eq. 9, except that $J$ has been replaced by $J\sqrt{2}$ in them. Also, they satisfy a relation as (10). It is desirable that for a particular time, namely $t_{\hat{H}_{\mu0}}$, we have
\begin{equation}
\hat{U}(t_{\hat{H}_{\mu}})|\mu-\nu_{0}^{a}\rangle=|\xi_{\mu0}^{a}\rangle,
\end{equation}
or equivalently
\begin{equation}
u_{\hat{H}_{\mu0},4}(t_{\hat{H}_{\mu0}})=u_{\hat{H}_{\mu0},2}(0)=1,
\end{equation}
which means that the state $|\mu-\nu_{0}^{a}\rangle$ has been transferred to $|\xi_{\mu0}^{a}\rangle$ perfectly at transfer time $t_{\hat{H}_{\mu0}}$ where is observed in Fig. 5. Now, after time $t_{\hat{H}_{\mu0}}$, if one of the following local operation is applied on the atomic states at the control part, i.e.
\begin{equation}
\begin{array}{c}
  \hat{\sigma}_{{Z}\mu_{2}}\hat{\sigma}_{Z\mu_{3}}|\xi_{\mu0}^{a}\rangle=|\xi_{\mu1}^{a}\rangle, \\\\
  \hat{\sigma}_{{Z}\mu_{1}}\hat{\sigma}_{Z\mu_{3}}|\xi_{\mu0}^{a}\rangle=|\xi_{\mu2}^{a}\rangle, \\\\
  \hat{\sigma}_{{Z}\mu_{1}}\hat{\sigma}_{Z\mu_{2}}|\xi_{\mu0}^{a}\rangle=|\xi_{\mu3}^{a}\rangle,
\end{array}
\end{equation}
then the dynamics of the system, in the switched subspace $\mathcal{H}_{\mu i}$, evolves the states $|\xi_{\mu i}^{a}\rangle$ to its relative states $|\mu-\nu_{i}^{a}\rangle$ after transferring time $t_{\hat{H}_{\mu i}}=t_{\hat{H}_{\mu0}}$, for each $i=1,2,3$.

Let us consider a hexagonal lattice as depicted in Fig. 6. Let $\mu$ be a vertex of the lattice. The three links connected to this vertex lie on the plane of lattice. On the links, there are three cavities which each of them doped with a two-level atom denoted by $\mu-\nu_{1}$, $\mu-\nu_{2}$ and $\mu-\nu_{3}$. Also in addition, there is other cavity with the same atom which connects to the vertex $\mu$ denoted by $\mu-\nu_{0}$ and does not lie in the plane of lattice. The vertex $\mu$ has the same structure as the control part of the Hamiltonian $\hat{H}_{\mu}$ and therefore, this vertex along with the atom-cavity systems: $\mu-\nu_{1}$, $\mu-\nu_{2}$, $\mu-\nu_{3}$ and $\mu-\nu_{0}$, form a system with the same structure as the Hamiltonian $\hat{H}_{\mu}$. Consequently, the Hamiltonian for the total lattice can be written as
\begin{equation}
\hat{H}=\sum_{\mu}\hat{H}_{\mu},
\end{equation}
where $\hat{H}_{\mu}$, in this way, is called as local Hamiltonian connecting each vertex to its neighboring links and through these links to the other vertices. It is convenient to choose the set of basis $\{|\mu-\nu_{i}^{b}\rangle,|\xi_{\mu i}^{b}\rangle\}$ with $b\in\{c,a\}$ and $i=0,1,2,3$, for the set of vertices $\mu$ of the hexagonal lattice, leading to a direct sum structure for the Hamiltonian as follows
\begin{equation}
\hat{H}=\bigoplus_{\mu}\hat{H}_{\mu0}\oplus\bigoplus_{\{\mu,\lambda\}\in E}\hat{H}_{\mu,\lambda},
\end{equation}
where $\hat{H}_{\mu0}$, which is the same as (20), stands for upload of a quantum state to hexagonal lattice or download from it. While $\hat{H}_{\mu,\lambda}$ with $\{\mu,\lambda\}\in E$, for which $E$ is the set of edges of hexagonal lattice, stands for transferring the uploaded quantum state from one vertex to the other adjacent vertex on the hexagonal lattice through the related invariant subspace
\begin{equation}
\mathcal{H}_{\mu,\lambda}=span\{|\xi_{\mu i}^{b}\rangle,|\mu-\nu_{i}^{b}\rangle,|\xi_{\lambda j}^{b}\rangle\},
\end{equation}
where $i,j\neq 0$, $i\neq j$ and $b\in\{c,a\}$.

The Hamiltonian $\hat{H}_{\mu,\lambda}$ which governs the time evolution of the system from vertex $\mu$ to its adjacent vertex $\lambda$ is as
\begin{equation}
\ \hat{H}_{\mu,\lambda} = \left(
          \begin{array}{cccccc}
          \Omega_{c} & G & 2J & 0 & 0 & 0\\
          G & \Omega_{c}-\Delta & 0 & 0 & 0 & 0\\
          2J & 0 & \Omega_{c} & G & 2J & 0\\
          0 & 0 & G & \Omega_{c}-\Delta & 0 & 0\\
          0 & 0 & 2J & 0 & \Omega_{c} & G \\
          0 & 0 & 0 & 0 & G & \Omega_{c}-\Delta\\
          \end{array}
          \right),
\end{equation}
where, it is the same as the $\hat{H}_{2}$ in Eq. 8 except that $J$ has been replaced with $\sqrt{2}J$. As the previous cases, perfect state transfer can occur in this subspace in such way that if we prepare the system in the atomic state $|\xi_{\mu i}^{a}\rangle$, we obtain the state $|\lambda_{\mu j}^{a}\rangle$ perfectly after transfer time $t_{\hat{H}_{\mu,\lambda}}$ as shown in Fig. 7 in the resonance regime.

So far, we have seen that by the method, it is possible to upload a quantum state from an arbitrary sender and transfer it to an arbitrary receiver using the two-dimensional hexagonal lattice. This method can be easily extend to three-dimensional version. If two or more hexagonal lattice plane which all of them are parallel and each of them connect to its neighboring lattice by some intermediate atom-cavity systems, the method can work for three-dimensional cases, Fig. 12. The routing of quantum states between the neighboring lattices is made by the Hamiltonian like the $\hat{H}_{\mu,\lambda}$ as unit of routing.
\section{Robustness to photon loss}
In the previous sections, it has been shown that the routing of quantum information in a quantum network takes place by tailing dynamics of the system in the invariant subspaces through the performing local operations on the related control parts of the network. In each switched invariant subspaces the field mode of cavities and the atoms are in resonance interaction, $\Delta=0$, with each other as discussed previously and shown in Fig. 2, 3, 5, 7. It is evident that in each invariant subspaces the field mode of cavities are considerably populated which in turn, implies that the related average number of photons in each invariant subspace can be considerably large. But, in the presence of interaction between the system and environment, it is evident from
\cite{dav, fon} that: the larger the average number of photons inside the cavity, the faster will the
coherence decay. Therefore, the lose of coherency of the system is unavoided. To remove
the effect of decoherence on the system as arisen as above, we should prevent the populating
of the field mode of the cavities. This situation can be achieved for nonzero values of
the detuning parameter, i.e. $\Delta\neq0$, as shown in Fig. 8, 9, 10, 11. As it is evident, the detuning parameter $\Delta$ controls amount of average number of photon in each invariant subspace so it can be choice in such a way that leads to a vanishing amount for the average number of photon in the field mode of cavities. Therefore, during
the process of routing of quantum information throughout the quantum network which takes place
slowly in comparison to the resonance cases, the field mode of the cavities only virtually excited. In another words, routing of quantum information in a network constructed from channels of coupled cavities, can be done by virtual photons, protecting against decoherence via cavity decay \cite{ogd, zhe, maj}.

\section{Conclusions}
We have presented a protocol for perfect routing of quantum information throughout the regular network of cavities. Perfect routing of quantum information depends on the existence of perfect state transfer invariant subspaces and possibility for tailing these subspaces to each other by applying preferably simple local operations yielding the switching of information between the subspaces. This protocol utilizes the cavity fields, as carrier of information, to couple two-level atoms, as sources of storage of information, in the channels of the network. Also, the protocol can work for entanglement transfer, as well as state transfer, between two arbitrary region of the network. The interaction could be mediated by the exchange of virtual photons rather than real photons by reducing the populations of cavity fields, avoiding cavity-induced loss. Therefore, the perfect routing of quantum information in cavity QED networks can be achieved without occurring efficient decoherence. This, in turn, represents an interesting and progressive step toward the realization of quantum communication in quantum computers.

\newpage
\vspace{1cm} \setcounter{section}{0}
 \setcounter{equation}{0}
 \renewcommand{\theequation}{\arabic{equation}}
{\Large{Appendix:}}\\
The $u_{\hat{H}_{1},i}(t)$s in Eq. 9:
\begin{eqnarray}
\begin{array}{c}
u_{\hat{H}_{1},1}(t)=\frac{A^{2}-(J\sqrt{2}+\Delta)^{2}}{8AG} e^{-i(\Omega_{c} + (J \sqrt{2}+ A-\Delta)/2)t}-\frac{A^{2}-(J\sqrt{2}+\Delta)^{2}}{8AG}e^{-i(\Omega_{c} + (J \sqrt{2} - A-\Delta)/2)t} \\\\
+\frac{B^{2}-(J\sqrt{2}-\Delta)^{2}}{8BG} e^{-i(\Omega_{c} - (J\sqrt{2} - B+\Delta)/2)t}-\frac{B^{2}-(J\sqrt{2}-\Delta)^{2}}{8BG} e^{-i(\Omega_{c} - (J\sqrt{2} + B+\Delta)/2)t}, \\\\
u_{\hat{H}_{1},2}(t)=\frac{A^{2}-(J\sqrt{2}+\Delta)^{2}}{8AG} e^{-i(\Omega_{c} + (J \sqrt{2}+ A-\Delta)/2)t}-\frac{A^{2}-(J\sqrt{2}+\Delta)^{2}}{8AG}e^{-i(\Omega_{c} + (J \sqrt{2} - A-\Delta)/2)t} \\\\
-\frac{B^{2}-(J\sqrt{2}-\Delta)^{2}}{8BG} e^{-i(\Omega_{c} - (J\sqrt{2} - B+\Delta)/2)t}+\frac{B^{2}-(J\sqrt{2}-\Delta)^{2}}{8BG} e^{-i(\Omega_{c} - (J\sqrt{2} + B+\Delta)/2)t}, \\\\
u_{\hat{H}_{1},3}(t)=\frac{A-J\sqrt{2}-\Delta}{4A}e^{-i(\Omega_{c} + (J\sqrt{2} + A-\Delta)/2)t}+\frac{A+J\sqrt{2}+\Delta}{4A}e^{-i(\Omega_{c} + (J\sqrt{2} - A-\Delta)/2)t}\\\\+\frac{B+J\sqrt{2}-\Delta}{4B}e^{-i(\Omega_{c} - (J\sqrt{2} - B+\Delta)/2)t}+\frac{B-J\sqrt{2}+\Delta}{4B}e^{-i(\Omega_{c} - (J\sqrt{2} + B+\Delta)/2)t}, \\\\
u_{\hat{H}_{1},4}(t)=\frac{A-J\sqrt{2}-\Delta}{4A}e^{-i(\Omega_{c} + (J\sqrt{2} + A-\Delta)/2)t}+\frac{A+J\sqrt{2}+\Delta}{4A}e^{-i(\Omega_{c} + (J\sqrt{2} - A-\Delta)/2)t}\\\\-\frac{B+J\sqrt{2}-\Delta}{4B}e^{-i(\Omega_{c} - (J\sqrt{2} - B+\Delta)/2)t}-\frac{B-J\sqrt{2}+\Delta}{4B}e^{-i(\Omega_{c} - (J\sqrt{2} + B+\Delta)/2)t},
\end{array}
\end{eqnarray}
in which $A = \sqrt {2 J^2 + 2\sqrt{2}J \Delta + \Delta^2 + 4 G^2}$ and $B = \sqrt {2 J^2 - 2\sqrt{2}J \Delta + \Delta^2 + 4 G^2}$.

The $u_{\hat{H}_{2},i}(t)$s in Eq. 12:
\begin{eqnarray}
\begin{array}{c}
u_{\hat{H}_{2},1}(t)=-\frac{C_{1}^{2}-\Delta^{2}}{8GC_{1}}e^{-i(\Omega_{c} - (C_{1}+\Delta)/2)t}+\frac{C_{1}^{2}-\Delta^{2}}{8GC_{1}}e^{-i(\Omega_{c} + (C_{1}-\Delta)/2)t}-\frac{C_{2}^2-(2J+\Delta)^2}{16GC_{2}}e^{-i(\Omega_{c} + J - (C_{2}+\Delta)/2)t}\\\\
+\frac{C_{2}^2-(2J+\Delta)^2}{16GC_{2}}e^{-i(\Omega_{c} + J + (C_{2}-\Delta)/2)t}-\frac{C_{3}^2-(2J-\Delta)^2}{16GC_{3}}e^{-i(\Omega_{c} - J- (C_{3}+\Delta)/2)t}+\frac{C_{3}^2-(2J-\Delta)^2}{16GC_{3}}e^{-i(\Omega_{c} - J + (C_{3}-\Delta)/2)t}\\\\
u_{\hat{H}_{2},2}(t)=\frac{C_{2}+2J+\Delta}{4\sqrt{2}C_{2}}e^{-i(\Omega_{c} + J - (C_{2}+\Delta)/2)t}+\frac{C_{2}-2J-\Delta}{4\sqrt{2}C_{2}}e^{-i(\Omega_{c} + J + (C_{2}-\Delta)/2)t}\\\\
-\frac{C_{2}-2J+\Delta}{4\sqrt{2}C_{2}}e^{-i(\Omega_{c} - J - (C_{3}+\Delta)/2)t}-\frac{C_{2}+2J-\Delta}{4\sqrt{2}C_{2}}e^{-i(\Omega_{c} - J + (C_{3}-\Delta)/2)t}\\\\
u_{\hat{H}_{2},3}(t)=\frac{C_{1}+\Delta}{4C_{1}}e^{-i(\Omega_{c} - (C_{1}+\Delta)/2)t}+\frac{C_{1}-\Delta}{4C_{1}}e^{-i(\Omega_{c} + (C_{1}-\Delta)/2)t}+\frac{C_{2}+2J+\Delta}{8C_{2}}e^{-i(\Omega_{c} + J - (C_{2}+\Delta)/2)t}\\\\
+\frac{C_{2}-2J-\Delta}{8C_{2}}e^{-i(\Omega_{c} + J + (C_{2}-\Delta)/2)t}+\frac{C_{3}-2J+\Delta}{8C_{3}}e^{-i(\Omega_{c} - J - (C_{3}+\Delta)/2)t}+\frac{F_{3}+2J-\Delta}{8C_{3}}e^{-i(\Omega_{c} - J + (C_{3}-\Delta)/2)t}\\\\
u_{\hat{H}_{2},4}(t)=\frac{C_{1}^{2}-\Delta^{2}}{8GC_{1}}e^{-i(\Omega_{c} - (C_{1}+\Delta)/2)t}-\frac{C_{1}^{2}-\Delta^{2}}{8GC_{1}}e^{-i(\Omega_{c} + (C_{1}-\Delta)/2)t}-\frac{C_{2}^2-(2J+\Delta)^2}{16GC_{2}}e^{-i(\Omega_{c} + J - (C_{2}+\Delta)/2)t}\\\\
+\frac{C_{2}^2-(2J+\Delta)^2}{16GC_{2}}e^{-i(\Omega_{c} + J + (C_{2}-\Delta)/2)t}-\frac{C_{3}^2-(2J-\Delta)^2}{16GC_{3}}e^{-i(\Omega_{c} - J- (C_{3}+\Delta)/2)t}+\frac{C_{3}^2-(2J-\Delta)^2}{16GC_{3}}e^{-i(\Omega_{c} - J + (C_{3}-\Delta)/2)t}\\\\
u_{\hat{H}_{2},5}(t)=-\frac{C_{2}^2-(2c+\Delta)^2}{8\sqrt{2}C_{2}G}e^{-i(\Omega_{c} + J - (C_{2}+\Delta)/2)t}+\frac{C_{2}^2-(2c+\Delta)^2}{8\sqrt{2}C_{2}G}e^{-i(\Omega_{c} + J + (C_{2}-\Delta)/2)t}\\\\
+\frac{C_{2}^2-(2c-\Delta)^2}{8\sqrt{2}C_{2}G}e^{-i(\Omega_{c} - J - (C_{3}+\Delta)/2)t}-\frac{C_{2}^2-(2c-\Delta)^2}{8\sqrt{2}C_{2}G}e^{-i(\Omega_{c} - J + (C_{3}-\Delta)/2)t}\\\\
u_{\hat{H}_{2},6}(t)=-\frac{C_{1}+\Delta}{4C_{1}}e^{-i(\Omega_{c} - (C_{1}+\Delta)/2)t}-\frac{C_{1}-\Delta}{4C_{1}}e^{-i(\Omega_{c} + (C_{1}-\Delta)/2)t}+\frac{C_{2}+2J+\Delta}{8C_{2}}e^{-i(\Omega_{c} + J - (C_{2}+\Delta)/2)t}\\\\
+\frac{C_{2}-2J-\Delta}{8C_{2}}e^{-i(\Omega_{c} + J + (C_{2}-\Delta)/2)t}+\frac{C_{3}-2J+\Delta}{8C_{3}}e^{-i(\Omega_{c} - J - (C_{3}+\Delta)/2)t}+\frac{F_{3}+2J-\Delta}{8C_{3}}e^{-i(\Omega_{c} - J + (C_{3}-\Delta)/2)t}\\\\
\end{array}
\end{eqnarray}
in which $C_{1}=\sqrt{\Delta^{2}+4G^{2}}$ and $C_{2}=\sqrt{4G^{2}+(2J+\Delta)^2}$ and $C_{3}=\sqrt{4G^{2}+(2J-\Delta)^2}$.

\newpage
\textbf{Figure Captions}
\itemize{}
\item Fig. 1. (a) 1D structure for routing of quantum information. Quasi-linear cavity system with quasi uniform  coupling of cavities, the black circles in the cavities are two-level atoms and two typical atom-cavity inside the dotted frame are  as control part of the network. (b) A linear network equivalent to quasi-linear one in which red circles play the role of the control part which switches the time evolution between invariant subspaces. The blue circles are the intermediate atom-cavity systems and the black circles are as sender and receiver ones.
\begin{figure}
\centering
\includegraphics[width=300 pt]{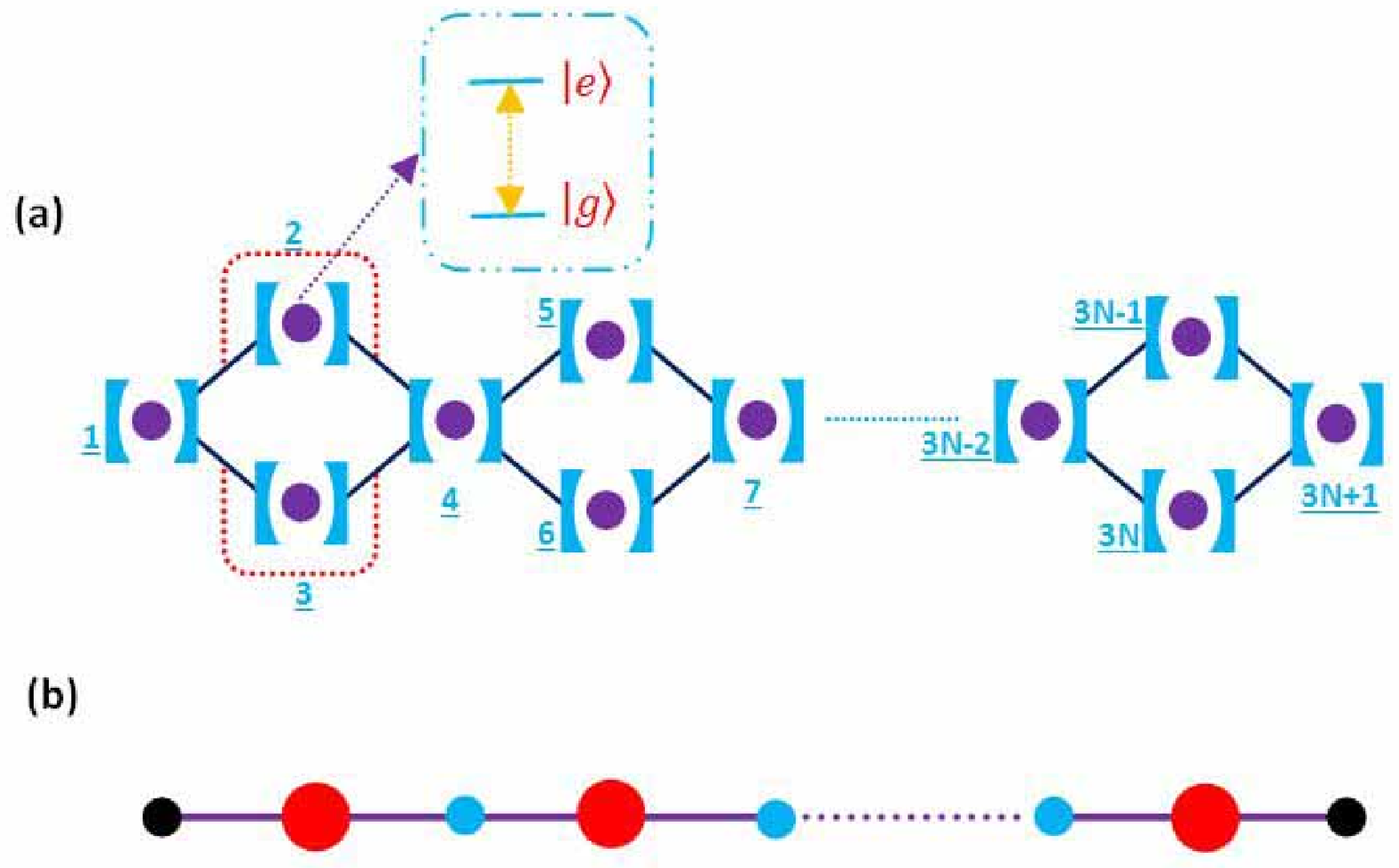}
\caption{} \label{Fig1}
\end{figure}
\newpage
\textbf{Figure Captions}
\item Fig. 2. Perfect state transfer in the invariant subspace $\mathcal{H}_{1}$ in (6), (or in the $\mathcal{H}_{N+1}$) for c=1, g=65 and $\Delta=0$ (in units of $\Omega_{c}$). The red curve represents
the population of the field mode of cavities, i.e. $F = |u_{\hat{H}_{1},1}(t)|^{2} + |u_{\hat{H}_{1},3}(t)|^{2}$. While $U_{1}=|u_{\hat{H}_{1},2}(t)|^{2}$ and $U_{2}=|u_{\hat{H}_{1},4}(t)|^{2}$ represented by
yellow and green curves respectively, are the populations of the related atoms. The expression $U_{2}=|u_{\hat{H}_{1},4}(t^{\ast})|^{2}$ = 1 with transfer time $t^{\ast}=2.2231$,
ensures perfect state transfer process in this invariant subspace.
\begin{figure}
\centering
\includegraphics[width=400 pt]{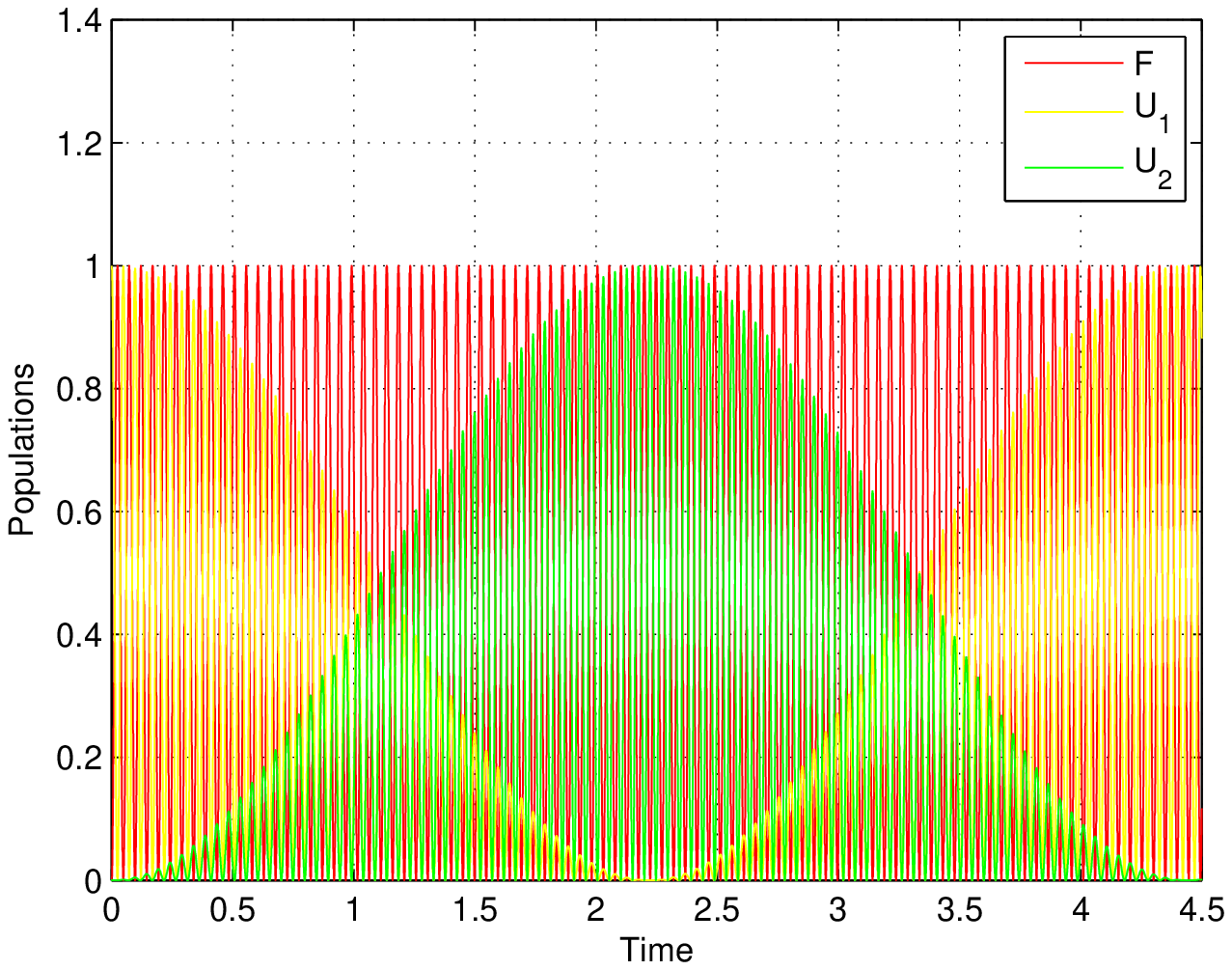}
\caption{} \label{Fig4}
\end{figure}
\newpage
\textbf{Figure Captions}
\item Fig. 3. Perfect state transfer in the invariant subspace $\mathcal{H}_{2}$ in (6), (or in the $\mathcal{H}_{3},..., \mathcal{H}_{N}$) for c=1, g=65 and $\Delta=0$ (in units of $\Omega_{c}$). The red curve represents
the population of the field mode of cavities, i.e. $F = |u_{\hat{H}_{2},1}(t)|^{2} + |u_{\hat{H}_{2},3}(t)|^{2}+|u_{\hat{H}_{2},5}(t)|^{2}$. $U_{1}=|u_{\hat{H}_{2},2}(t)|^{2}$, $U_{2}=|u_{\hat{H}_{2},4}(t)|^{2}$ and $U_{3}=|u_{\hat{H}_{2},6}(t)|^{2}$ represented by
yellow, blue and green curves respectively, are the populations of the related atoms. In this subspace, the statement  $U_{3}=|u_{\hat{H}_{2},6}(t^{\ast})|^{2}$ = 1 with transfer time $t^{\ast}=3.1410$, certifies the existence of perfect state transfer process in this invariant subspace.
\begin{figure}
\centering
\includegraphics[width=400 pt]{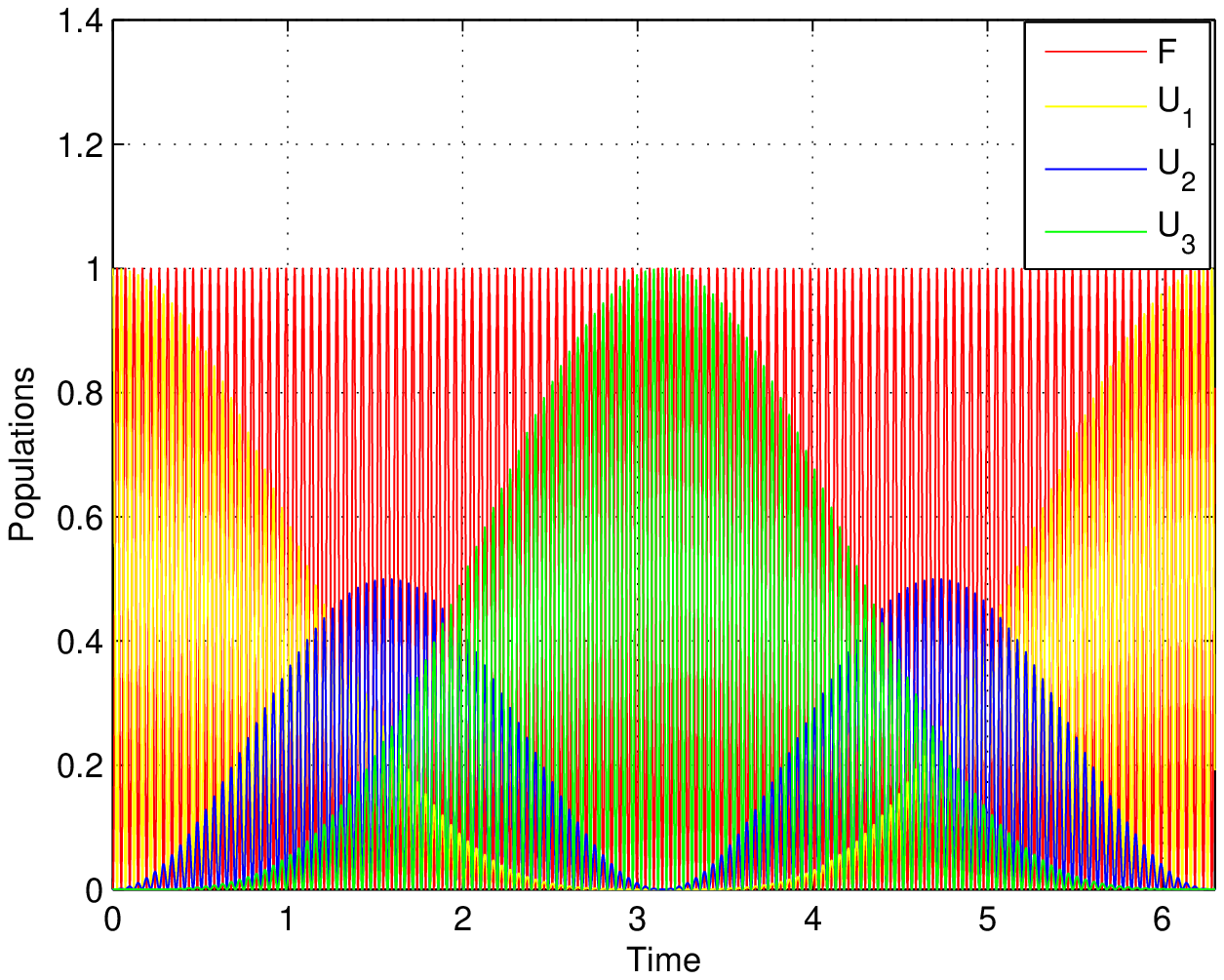}
\caption{} \label{Fig5}
\end{figure}
\newpage
\textbf{Figure Captions}
\item Fig. 4. The 3D switch structure. The system of four atom-cavity contained in the dotted-dashed frame called control part, the atom-cavity systems outside the frame denoted by 1, 2, 3, connect the switch to the other parts of the network. The other atom-cavity system outside the frame denoted by 0, devoted for uploading or downloading the quantum states.
\begin{figure}
\centering
\includegraphics[width=300 pt]{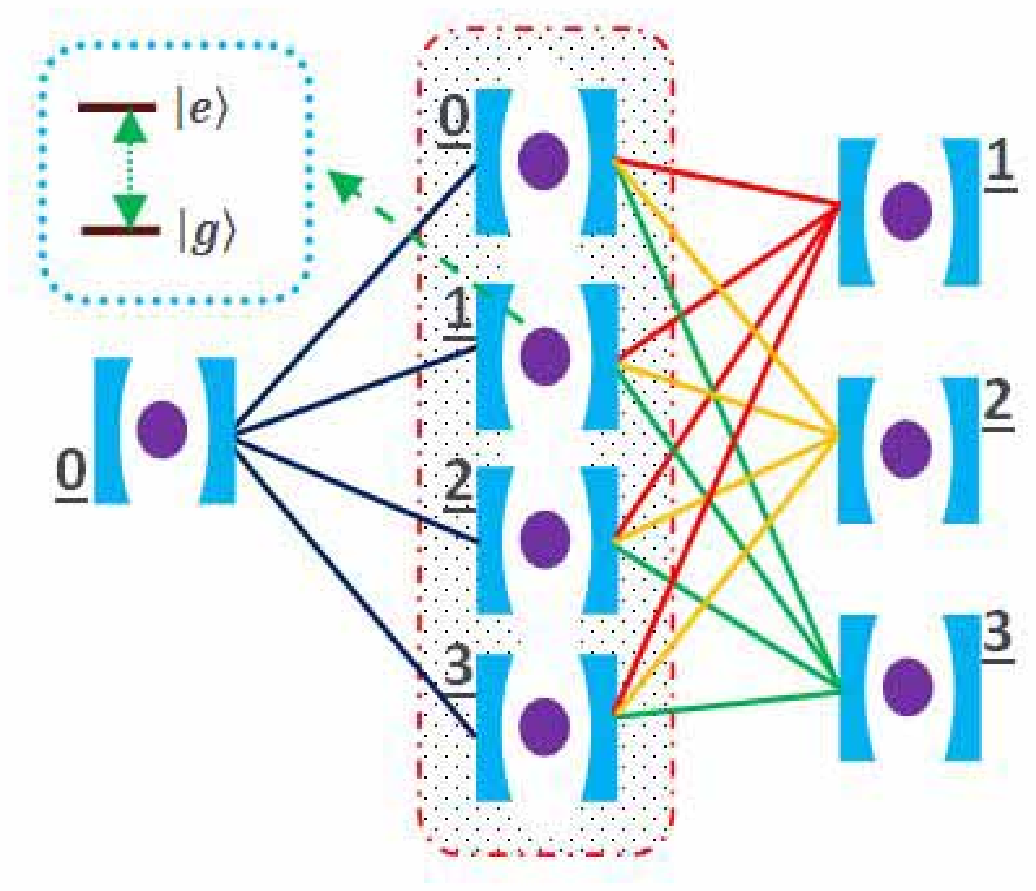}
\caption{} \label{Fig4}
\end{figure}
\newpage
\textbf{Figure Captions}
\item Fig. 5. Perfect state transfer in the invariant subspace $\mathcal{H}_{\mu0}$ in (20) (or in the $\mathcal{H}_{\mu i}, i=1,2,3$) for c=1, g=65 and $\Delta=0$ (in units of $\Omega_{c}$). The red curve represents
the population of the field mode of cavities, i.e. $F = |u_{\hat{H}_{\mu0},1}(t)|^{2} + |u_{\hat{H}_{\mu0},3}(t)|^{2}$. While $U_{1}=|u_{\hat{H}_{\mu0},2}(t)|^{2}$ and $U_{2}=|u_{\hat{H}_{\mu0},4}(t)|^{2}$ represented by
yellow and green curves respectively, are the populations of the related atoms. The expression $U_{2}=|u_{\hat{H}_{\mu0},4}(t^{\ast})|^{2}$ = 1 with transfer time $t^{\ast}=1.5948$,
ensures perfect state transfer process in this invariant subspace.
\begin{figure}
\centering
\includegraphics[width=400 pt]{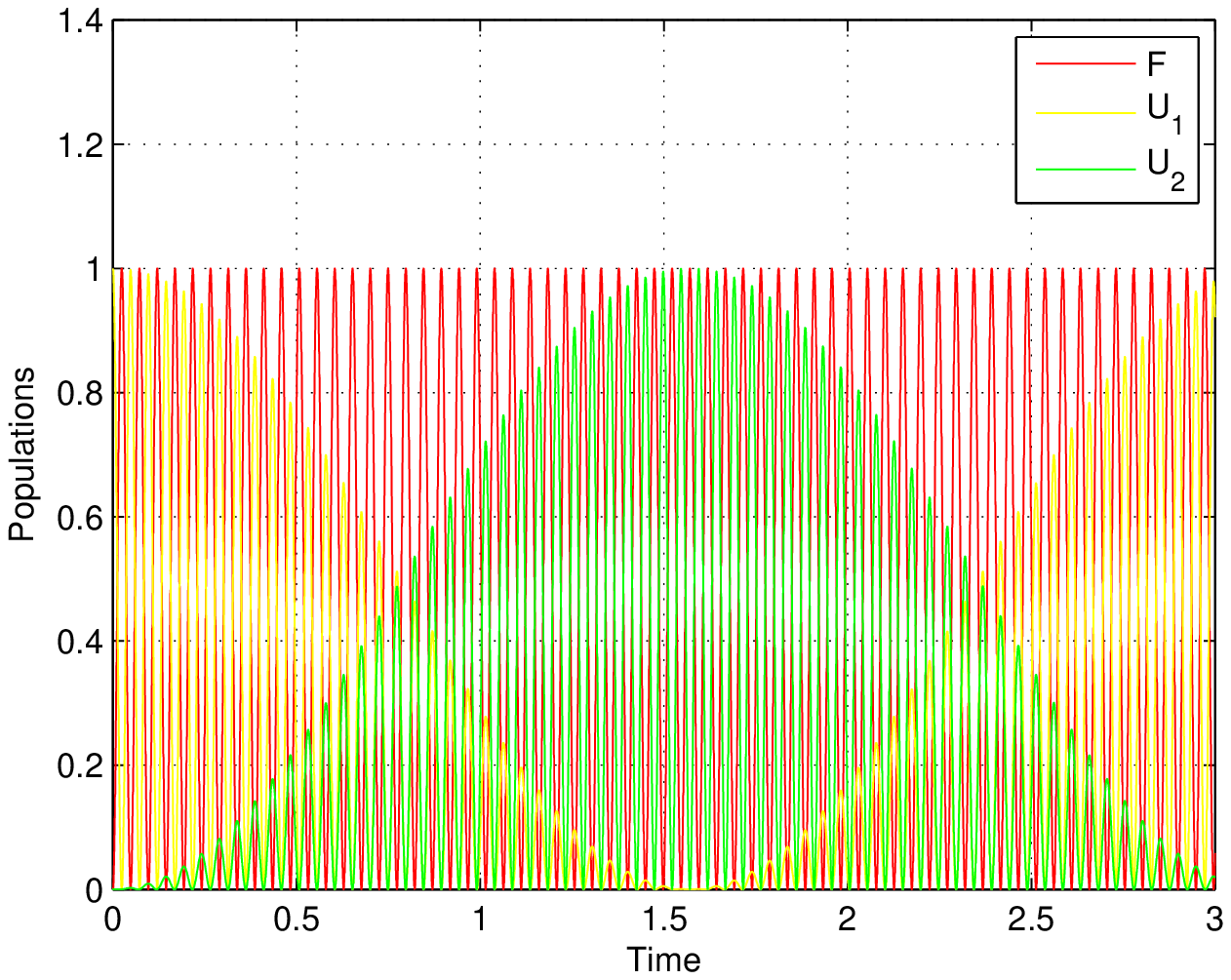}
\caption{} \label{Fig5}
\end{figure}
\newpage
\textbf{Figure Captions}
\item Fig. 6. Routing on a hexagonal lattice. The edges of the hexagonal lattice correspond to perfect state transfer invariant subspaces described by the Hamiltonian in (28).  Uploading the quantum states to the hexagonal lattice and downloading from it are performed in black circles by the $\hat{H}_{\mu0}$ in (20). The 3D switches are used to switch the quantum states in three directions on hexagonal lattice and also to the upload and download black circles. A quantum state can be uploaded from A to the hexagonal lattice and routed along the dashed path and finally downloaded to B perfectly.
\begin{figure}
\centering
\includegraphics[width=300 pt]{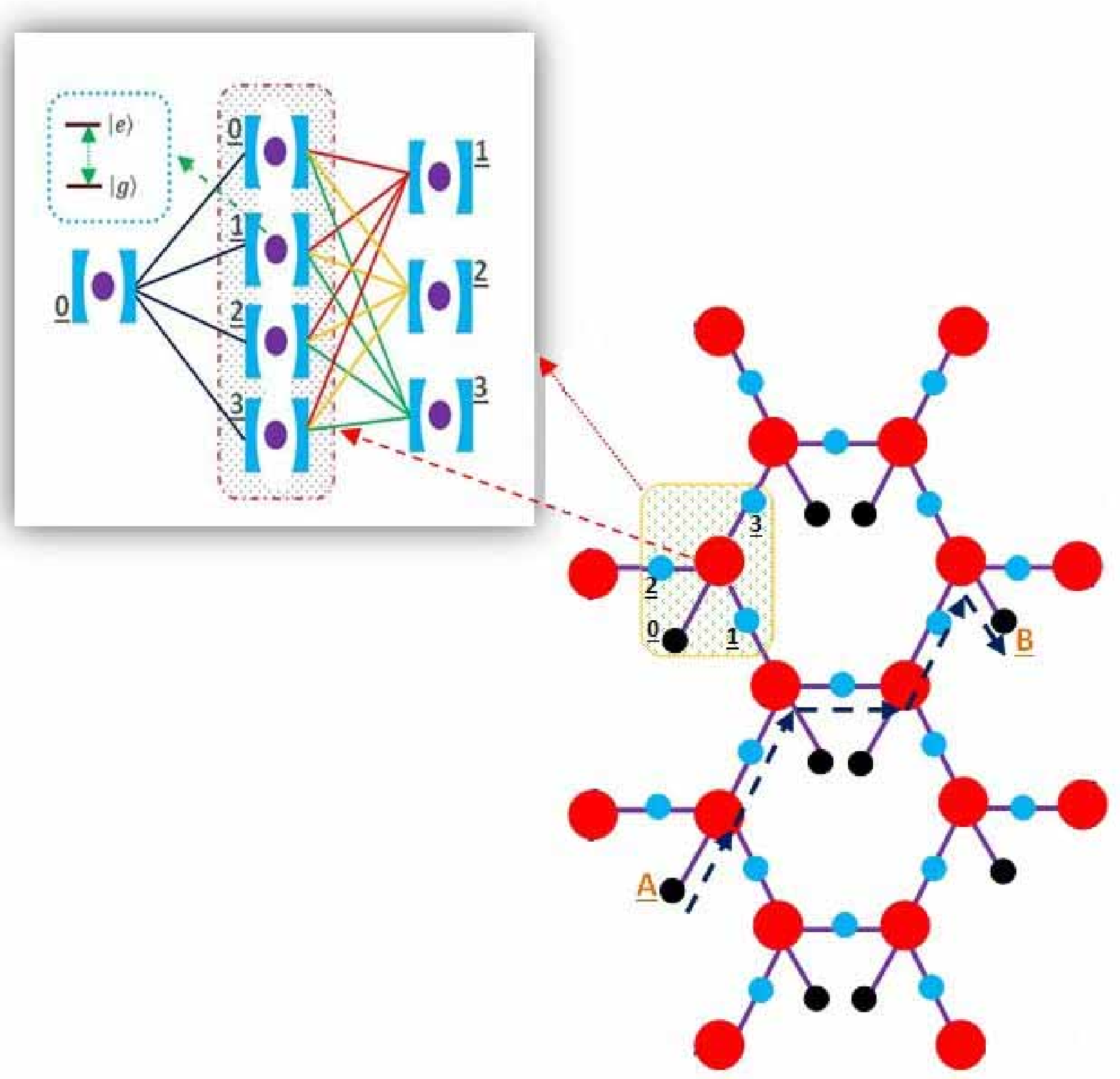}
\caption{} \label{Fig3}
\end{figure}
\newpage
\textbf{Figure Captions}
\item Fig. 7. Perfect state transfer in the invariant subspace $\mathcal{H}_{\mu,\lambda}$ in (28) for c=1, g=65 and $\Delta=0$ (in units of $\Omega_{c}$). The red curve represents
the population of the field mode of cavities, i.e. $F = |u_{\hat{H}_{\mu,\lambda},1}(t)|^{2} + |u_{\hat{H}_{\mu,\lambda},3}(t)|^{2}+|u_{\hat{H}_{\mu,\lambda},5}(t)|^{2}$. $U_{1}=|u_{\hat{H}_{\mu,\lambda},2}(t)|^{2}$, $U_{2}=|u_{\hat{H}_{\mu,\lambda},4}(t)|^{2}$ and $U_{3}=|u_{\hat{H}_{\mu,\lambda},6}(t)|^{2}$ represented by
yellow, blue and green curves respectively, are the populations of the related atoms. In this subspace, the statement  $U_{3}=|u_{\hat{H}_{\mu,\lambda},6}(t^{\ast})|^{2}$ = 1 with transfer time $t^{\ast}=2.2230$, certifies the existence of perfect state transfer process in this invariant subspace.
\begin{figure}
\centering
\includegraphics[width=400 pt]{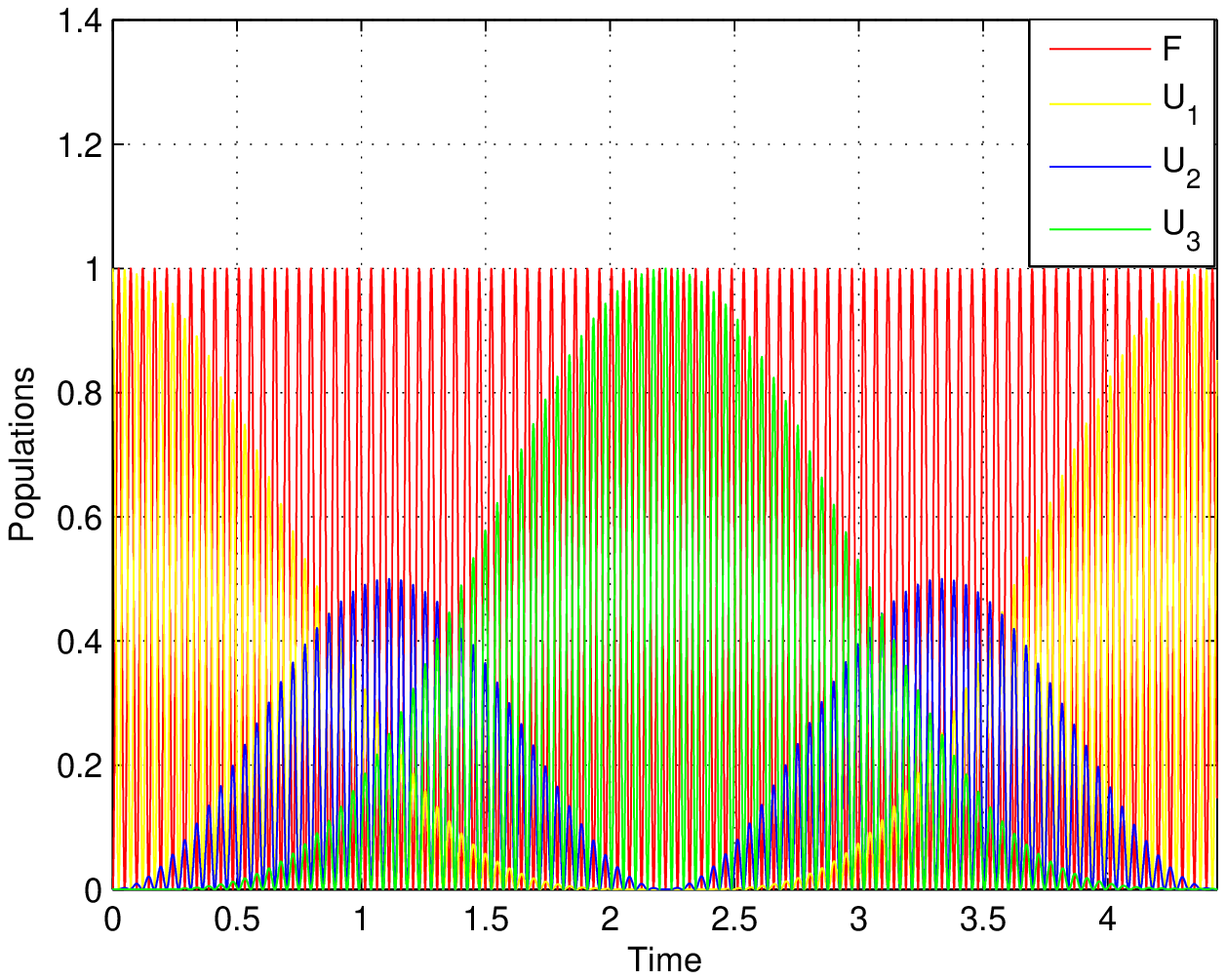}
\caption{} \label{Fig5}
\end{figure}
\newpage
\textbf{Figure Captions}
\item Fig. 8. Perfect state transfer in the invariant subspace $\mathcal{H}_{1}$ in (6) (or in the $\mathcal{H}_{N+1}$), at the nonresonance regime with c=1, g=65, $\Delta=-1000$ (in units of $\Omega_{c}$). The red curve shows a vanishing population for the field mode of cavities in this invariant subspace. $U_{1}=|u_{\hat{H}_{1},2}(t)|^{2}$ and $U_{2}=|u_{\hat{H}_{1},4}(t)|^{2}$ show the populations for the atoms represented by
yellow and green curves respectively. $U_{2}=|u_{\hat{H}_{1},4}(t^{\ast})|^{2}$ = 1 with transfer time $t^{\ast}=266.5300$, indicates the occurrence of perfect state transfer in this invariant subspace.
\begin{figure}
\centering
\includegraphics[width=400 pt]{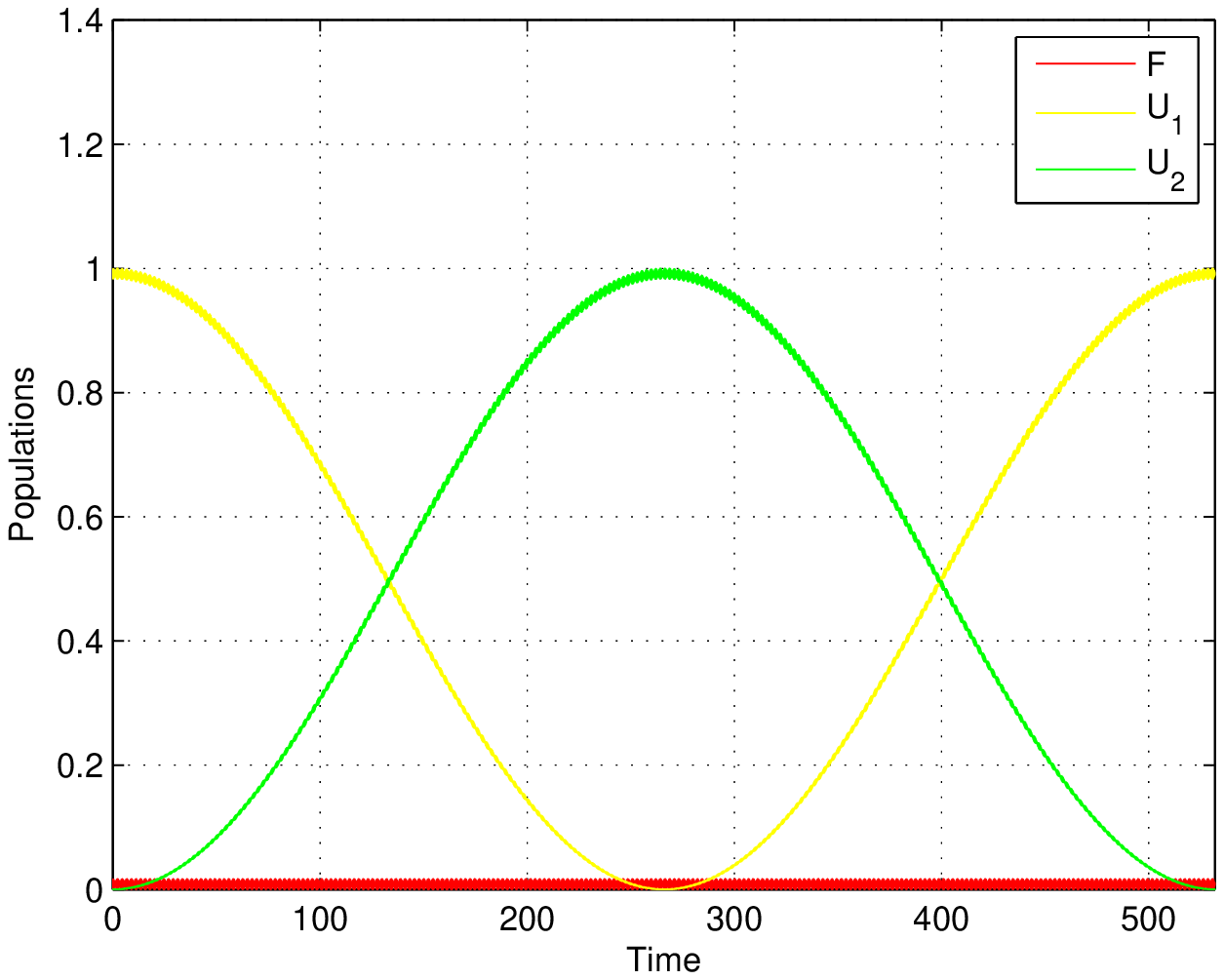}
\caption{} \label{Fig5}
\end{figure}
\newpage
\textbf{Figure Captions}
\item Fig. 9.  Perfect state transfer in the invariant subspace $\mathcal{H}_{2}$ in (6) (or in the $\mathcal{H}_{3},..., \mathcal{H}_{N}$), at nonresonance case with c=1, g=65 and $\Delta=-1000$ (in units of $\Omega_{c}$). The red curve shows
a vanishing population for the field mode of cavities and $U_{1}=|u_{\hat{H}_{2},2}(t)|^{2}$, $U_{2}=|u_{\hat{H}_{2},4}(t)|^{2}$ , $U_{3}=|u_{\hat{H}_{2},6}(t)|^{2}$ represented by
yellow, blue and green curves respectively, are the populations of the related atoms. In this subspace, the statement  $U_{3}=|u_{\hat{H}_{2},6}(t^{\ast})|^{2}$ = 1 with transfer time $t^{\ast}=376.9670$, certifies the existence of perfect state transfer process in this invariant subspace.
\begin{figure}
\centering
\includegraphics[width=400 pt]{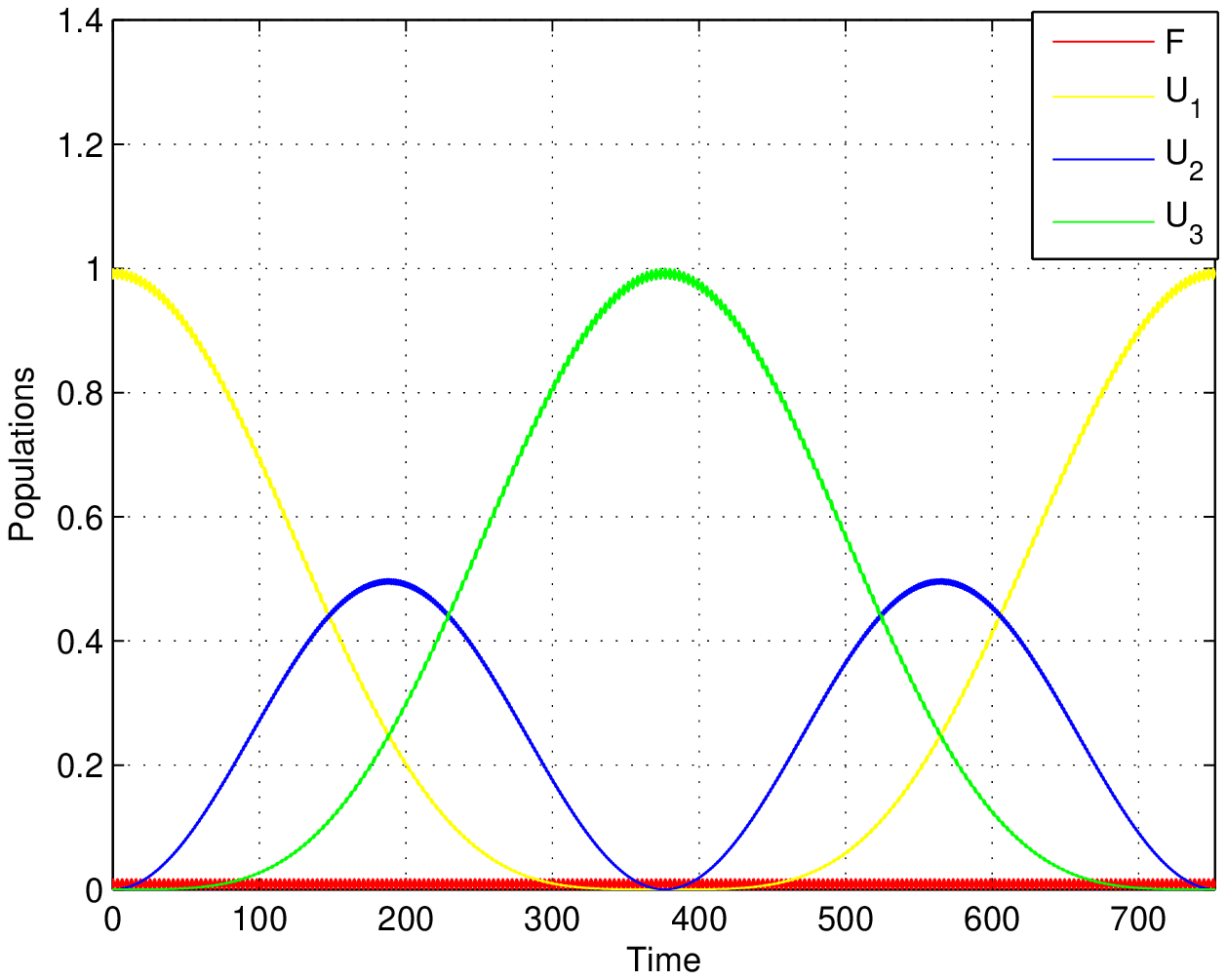}
\caption{} \label{Fig5}
\end{figure}
\newpage
\textbf{Figure Captions}
\item Fig. 10. Perfect state transfer in the invariant subspace $\mathcal{H}_{\mu0}$ in (20) (or in the $\mathcal{H}_{\mu i}, i=1,2,3$), at nonresonance case with c=1, g=65 and $\Delta=-1000$ (in units of $\Omega_{c}$). The red curve represents a vanishing population for the field mode of cavities and  $U_{1}=|u_{\hat{H}_{\mu0},2}(t)|^{2}$ and $U_{2}=|u_{\hat{H}_{\mu0},4}(t)|^{2}$ represented by yellow and green curves respectively, are the populations of the related atoms. The expression $U_{2}=|u_{\hat{H}_{\mu0},4}(t^{\ast})|^{2}$ = 1 with transfer time $t^{\ast}=188.4710$, ensures perfect state transfer process in this invariant subspace.
\begin{figure}
\centering
\includegraphics[width=400 pt]{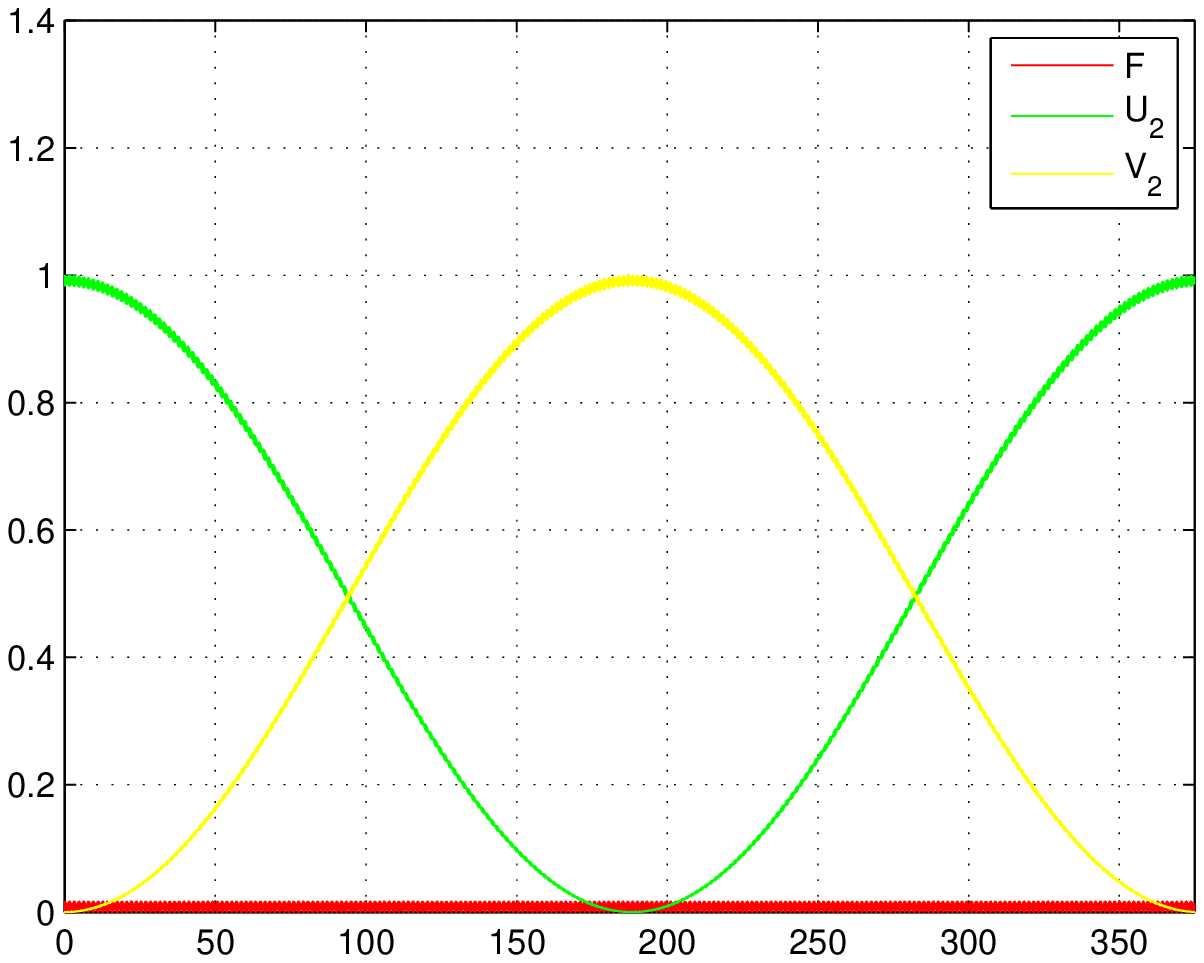}
\caption{} \label{Fig5}
\end{figure}
\newpage
\textbf{Figure Captions}
\item Fig. 11. Perfect state transfer in the invariant subspace $\mathcal{H}_{\mu,\lambda}$ in (28) for for the nonresonance case with c=1, g=65 and $\Delta=-1000$ (in units of $\Omega_{c}$). The red curve represents
a vanishing population of the field mode of cavities. $U_{1}=|u_{\hat{H}_{\mu,\lambda},2}(t)|^{2}$, $U_{2}=|u_{\hat{H}_{\mu,\lambda},4}(t)|^{2}$ and $U_{3}=|u_{\hat{H}_{\mu,\lambda},6}(t)|^{2}$ represented by
yellow, blue and green curves respectively, are the populations of the related atoms. In this subspace, the statement  $U_{3}=|u_{\hat{H}_{\mu,\lambda},6}(t^{\ast})|^{2}$ = 1 with transfer time $t^{\ast}=266.5580$, certifies the existence of perfect state transfer process in this invariant subspace.
\begin{figure}
\centering
\includegraphics[width=400 pt]{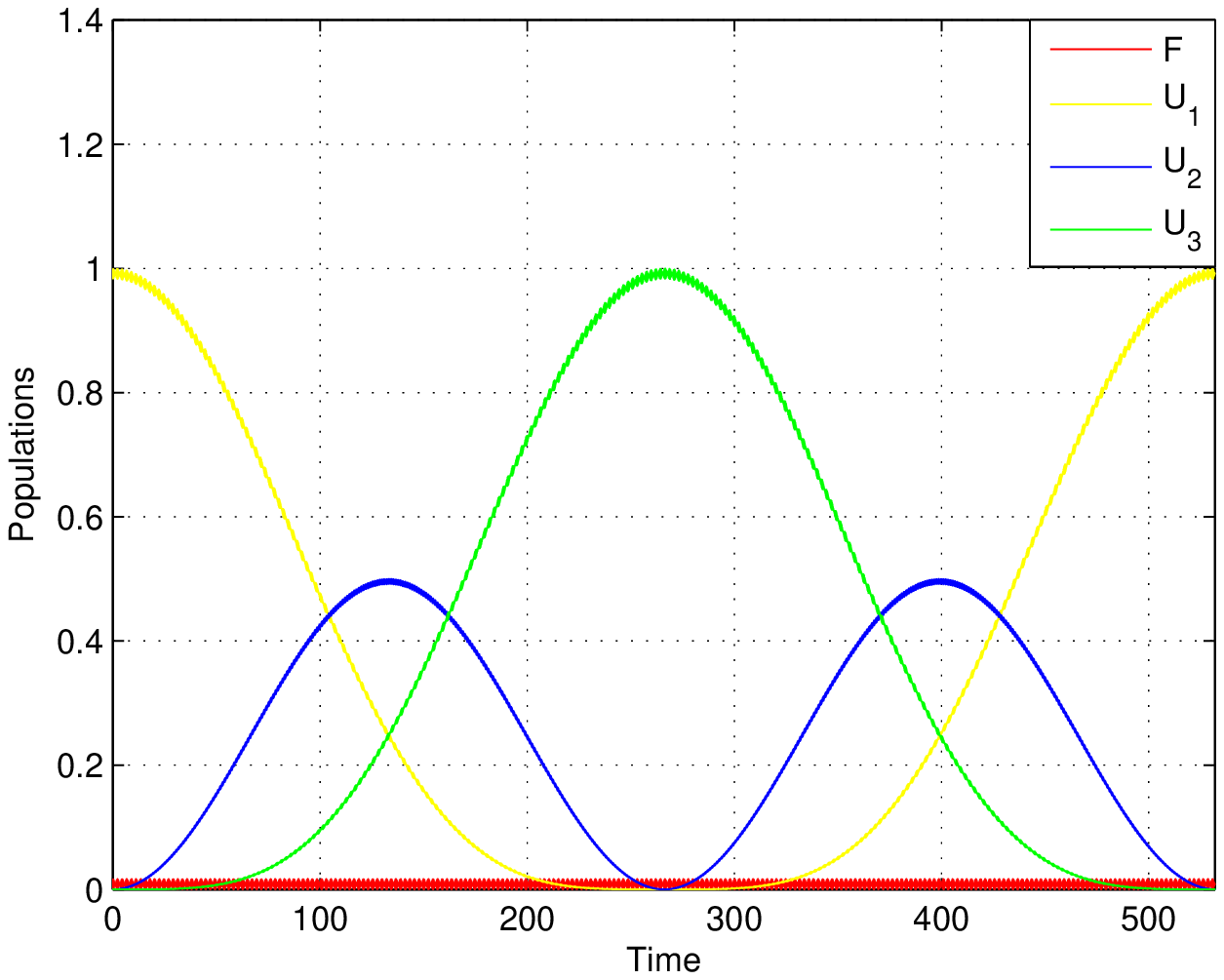}
\caption{} \label{Fig5}
\end{figure}
\newpage
\textbf{Figure Captions}
\item Fig. 12. 3D Perfect quantum routing utilizes two parallel hexagonal lattices.
\begin{figure}
\centering
\includegraphics[width=400 pt]{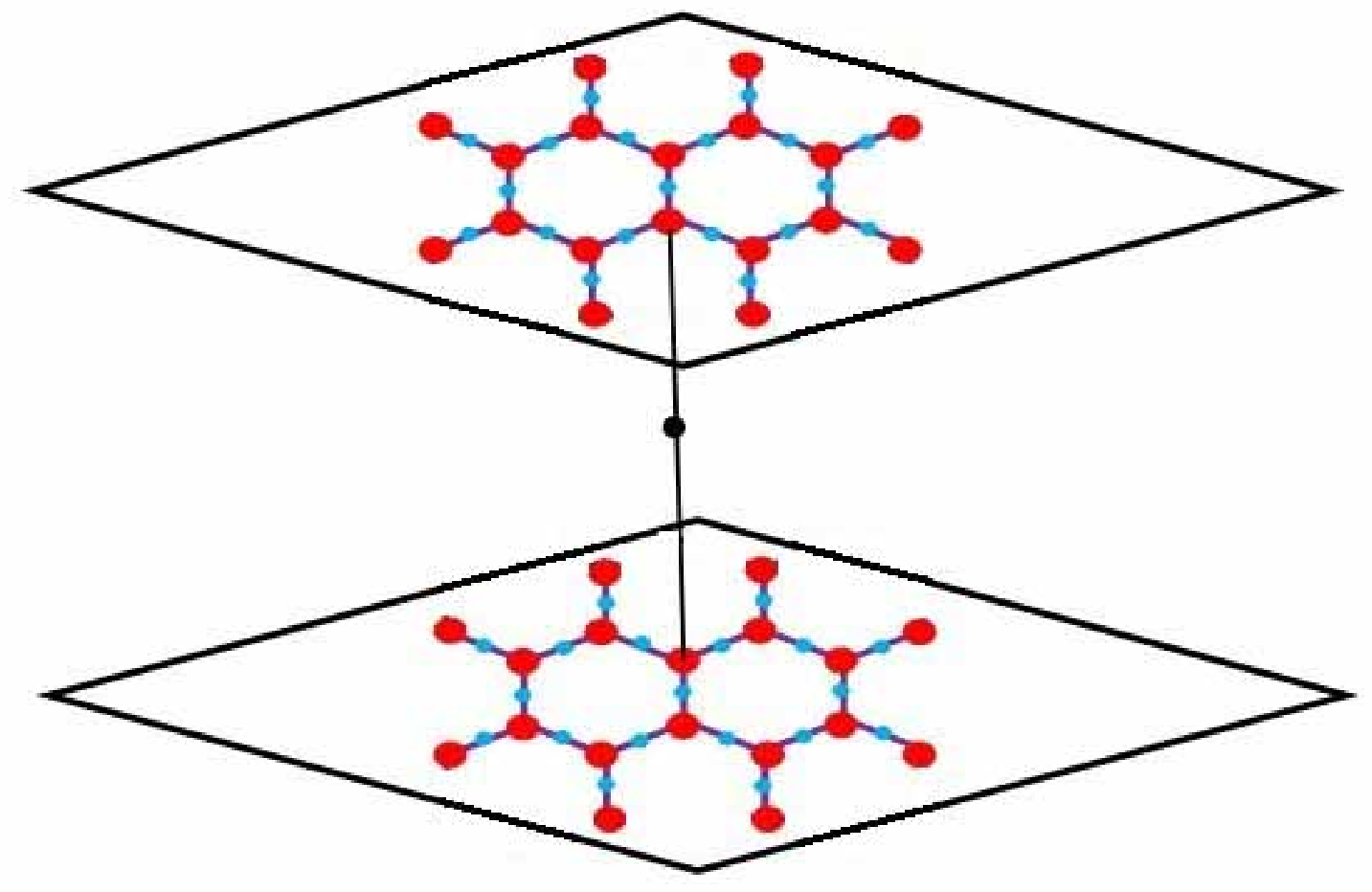}
\caption{} \label{Fig5}
\end{figure}

\end{document}